\documentclass[fleqn,usenatbib]{mnras}

\usepackage{newtxtext,newtxmath}

\usepackage[T1]{fontenc}
\usepackage{ae,aecompl}

\DeclareRobustCommand{\VAN}[3]{#2}
\let\VANthebibliography\thebibliography
\def\thebibliography{\DeclareRobustCommand{\VAN}[3]{##3}\VANthebibliography}

\usepackage{graphicx}	
\usepackage{amsmath}	
\usepackage{bm}

\title[A black hole in NGC 1850]{A black hole detected in the young massive LMC cluster NGC 1850}

\author[S. Saracino et al.]{S. Saracino$^{1}$\thanks{E-mail: s.saracino@ljmu.ac.uk},
S. Kamann$^{1}$,
M. G. Guarcello$^{2}$,
C. Usher$^{3}$,
N. Bastian$^{4,5,1}$,  
I. Cabrera-Ziri$^{6}$,\newauthor
M. Gieles$^{7,8}$,
S. Dreizler$^{9}$,
G. S. Da Costa$^{10}$,
T.-O. Husser$^{9}$,
V. H\'{e}nault-Brunet$^{11}$\\
\\
$^{1}$ Astrophysics Research Institute, Liverpool John Moores University, 146 Brownlow Hill, Liverpool L3 5RF, UK\\
$^{2}$ Osservatorio Astronomico di Palermo, Piazza del Parlamento 1, I-90134, Palermo, Italy\\
$^{3}$ Department of Astronomy, Oskar Klein Centre, Stockholm University, AlbaNova University Center, SE-106 91 Stockholm, Sweden\\
$^{4}$ Donostia International Physics Center (DIPC), Paseo Manuel de Lardizabal, 4, 20018, Donostia-San Sebasti\'an, Guipuzkoa, Spain\\
$^{5}$ IKERBASQUE, Basque Foundation for Science, 48013, Bilbao, Spain \\
$^{6}$ Astronomisches Rechen-Institut, Zentrum f\"ur Astronomie der Universit\"at Heidelberg, M\"onchhofstra{\ss}e 12-14, D-69120 Heidelberg, Germany\\
$^{7}$ ICREA, Pg. Llu\'{i}s Companys 23, E08010 Barcelona, Spain\\
$^{8}$ Institut de Ci\`{e}ncies del Cosmos (ICCUB), Universitat de Barcelona (IEEC-UB), Mart\'{i} i Franqu\`{e}s 1, E08028 Barcelona, Spain\\
$^{9}$ Institute for Astrophysics, Georg-August-University G\"ottingen, Friedrich-Hund-Platz 1, D-37077 G\"ottingen, Germany\\
$^{10}$ Research School of Astronomy and Astrophysics, Australian National University, Canberra, ACT 0200, Australia\\
$^{11}$ Department of Astronomy and Physics, Saint Mary’s University, 923 Robie Street, Halifax, NS B3H 3C3, Canada\\
}

\date{Accepted XXX. Received YYY; in original form ZZZ}
\pubyear{2021}

\begin{document}
\label{firstpage}
\pagerange{\pageref{firstpage}--\pageref{lastpage}}
\maketitle
\begin{abstract}
We report the detection of a black hole (NGC 1850 BH1) in the $\sim$100 Myr-old massive cluster NGC~1850 in the Large Magellanic Cloud. It is in a binary system with a main-sequence turn-off star (4.9 $\pm$ 0.4 M${_\odot}$), which is starting to fill its Roche Lobe and becoming distorted. Using 17 epochs of VLT/MUSE observations we detected radial velocity variations exceeding 300 km/s associated to the target star, linked to the ellipsoidal variations measured by OGLE-IV in the optical bands. Under the assumption of a semi-detached system, the simultaneous modelling of radial velocity and light curves constraints the orbital inclination of the binary to ($38 \pm 2$)$^{\circ}$, resulting in a true mass of the unseen companion of $11.1_{-2.4}^{+2.1}$ $M_{\odot}$. This represents the first direct dynamical detection of a black hole in a young massive cluster, opening up the possibility of studying the initial mass function and the early dynamical evolution of such compact objects in high-density environments.
\end{abstract}


\section{Introduction}
The tremendous number (50 in GWTC-2) of gravitational waves (GWs) detected by LIGO since 2015 \citep{Abbott2016b,Abbott2016a} gives a sense of how urgent and important it is to characterize the properties of their progenitors, i.e., dense compact objects like Neutron Stars (NSs) and Black Holes (BHs), and to study both their formation and evolutionary channels. Moreover, the recent discovery of a $\sim$150~M${_\odot}$ BH (the first secure detection of an intermediate-mass black hole, \citealt{Abbott2020a,Abbott2020b}) as the coalescence product of two very massive BHs (60 M${_\odot}$ and 85 M${_\odot}$, respectively) has challenged our understanding of stellar evolution in massive stars \citep{Vink2021}, moving the focus to high-density environments like massive stellar clusters, where merger cascades are most likely to happen (see the recent review by \citealt{Gerosa2021}). 
BHs, however, are elusive objects, and apart from GW emission of coalescing binary BHs, we have only two ways to detect them: indirectly, via the radio, X-ray or gamma ray emissions of matter accreting onto them, or directly, by studying the orbital motion of a visible companion orbiting around it in a binary system. Over the past decades, there have been numerous indirect detections of BHs as members of binaries with luminous companions, and many of these binary candidates have been identified through X-ray and radio observations of accreting systems (see \citealt{Cowley1992} as an example). The first BHs in old globular clusters (GCs) were also detected in this way \citep[e.g.][]{2007Natur.445..183M,2012Natur.490...71S, 2013ApJ...777...69C, 2015MNRAS.453.3918M}, challenging the classical idea that most BHs are expected to be ejected from the cluster on relatively short timescales ($< 10^9$ yr, \citealt{Kulkarni1993}). A few direct dynamical detections of non-interacting BHs have been made so far in star clusters (in the $\sim$12 Gyr old NGC 3201, \citealt{giesers2018,giesers2019}) due to observational limitations (i.e. the need for high spatial and spectral resolution in crowded fields and multiple epochs), hence we know very little about the initial mass distribution of BHs and their early dynamical evolution.
In addition to these signals of individual BHs, other observables may point at populations of stellar BHs in star clusters, such as the evolution of core radius with age \citep{Mackey2007,Mackey2008}, the absence of mass segregation of stars (\citealt{Peuten2016,Alessandrini2016,Weatherford2020}), the central mass-to-light ratio of Omega Cen (\citealt{Zocchi2019,Baumgardt2019}), the core over half-light radius (\citealt{Askar2018,Kremer2020}) and the presence of tidal tails \citep{Gieles2021}.
Young massive clusters ($<$ a few Gyrs) are the best places to look for shedding new light on the field. Indeed, the detection of BHs can provide crucial constraints on their retention fraction after supernova natal kicks (before significant dynamical evolution and dynamical ejections have taken place): a major uncertainty in GC models. However, to the best of our knowledge, no claims have been made so far about a direct dynamical detection of a BH in these objects.

We are currently conducting a systematic search for stellar-mass BHs in two young massive stellar clusters in the Large Magellanic Cloud by exploiting multi-epoch MUSE (Multi Unit Spectroscopic Explorer, \citealt{muse}) observations taken at the ESO Very Large Telescope, through the monitoring of radial velocity variations. This approach is highly sensitive to the detection of stellar companions of massive objects. Here we present the first outcome of the survey: the discovery of the first BH in NGC 1850, a massive (M~$\sim10^{5}$M${_\odot}$, \citealt{mass2005}) $\sim$100 Myr-old cluster. This is also one of the rare cases where the true BH mass can be estimated, as the inclination of the binary is well constrained from photometric light curves provided by OGLE (Optical Gravitational Lensing Experiment, \citealt{OGLE1992}), given an assumed configuration for the system.

\section{Observations and data reduction}\label{sec:obs}
We observed NGC 1850 with MUSE in wide-field mode (WFM, programs: 0102.D-0268 and 106.216T.001, PI: Bastian), taking advantage of the adaptive optics (AO) module, which provides a substantial improvement in terms of the spatial resolution of the images. In WFM, MUSE covers a field of view (FOV) of $1\times1$~arcmin at a spatial sampling of 0.2~arcsec. Each spaxel records a spectrum from 470 to 930~nm at a (nearly) constant FWHM of 0.25~nm, corresponding to a spectral resolution between 1\,700 at the blue end and 3\,500 at the red end of the spectral range. As demonstrated by \citet{kamann2016} using telluric absorption features, the wavelength calibration of MUSE is stable to $1\,{\rm km\,s^{-1}}$, both across the FOV and in between observations.

The observations consist of two pointings, separated by about 50~arcsec. One is centered on the cluster core, the other samples a slightly outer region (we will refer to them as center and outer pointings, hereafter). These data span a time baseline of almost 2 years, with a time sampling between individual epochs ranging from 1 hour to several months. This configuration ensures our sensitivity to binaries over a wide range of orbital periods within the cluster. Although a detailed description of the binary content of NGC~1850 will be presented in a forthcoming paper, we mention here that for every single bright source within either MUSE FOV we have a sample of 16 extracted spectra with good signal to noise (S/N), which goes up to 32 for stars in the overlapping region between center and outer pointings.

We used the standard ESO MUSE pipeline to reduce the MUSE raw data (ESO Reflex, \citealt{pipeline}), while the extraction of individual stellar spectra was performed with the latest version of \textsc{PampelMuse}, \citep{Kamann2013}, based on a PSF-fitting technique using the combined spatial and spectral information.
For a proper extraction of the spectra, high spatial resolution photometry is needed as a reference. We took advantage of archival \textit{Hubble} space telescope (HST) observations of NGC~1850, taken with the WFC3 camera during programmes 14069 (PI: Bastian) and 14174 (PI: Goudfrooij). The data, which consist of images in ultraviolet to optical filters, were analysed using a standard PSF-fitting technique within the photometric software \textsc{DOLPHOT} \citep{dolphin2016}. The magnitudes of bright stars, saturated in almost all the long exposures, were recovered using the shortest exposure (7s) in F814W. This catalog was then used as a reference for the extraction process within \textsc{PampelMuse}.

\section{Photometric and Spectroscopic Analysis}
Binary stars in GCs are expected to be in tight orbits in order to survive in such dense environments. This means that even in high spatial resolution images (e.g. HST), the two components should not be resolved. One possibility to detect these compact sources is via a systematic search for radial velocity variations, reflecting the precise orbital motion of one star around the other. Except for the case where both stars have similar brightness, the extracted spectrum will be always dominated by one of the stars. This approach is extremely powerful as it enables to distinguish between luminous and dark companions. While a few interesting targets have been identified in this way in NGC~1850, we here devote our attention on a specific target star in the sample, showing radial velocity variations exceeding 300 km/s, a clear footprint of an underlying binary system and a massive (likely dark) companion. 

\begin{figure*}
    \centering
	\includegraphics[width=1\textwidth]{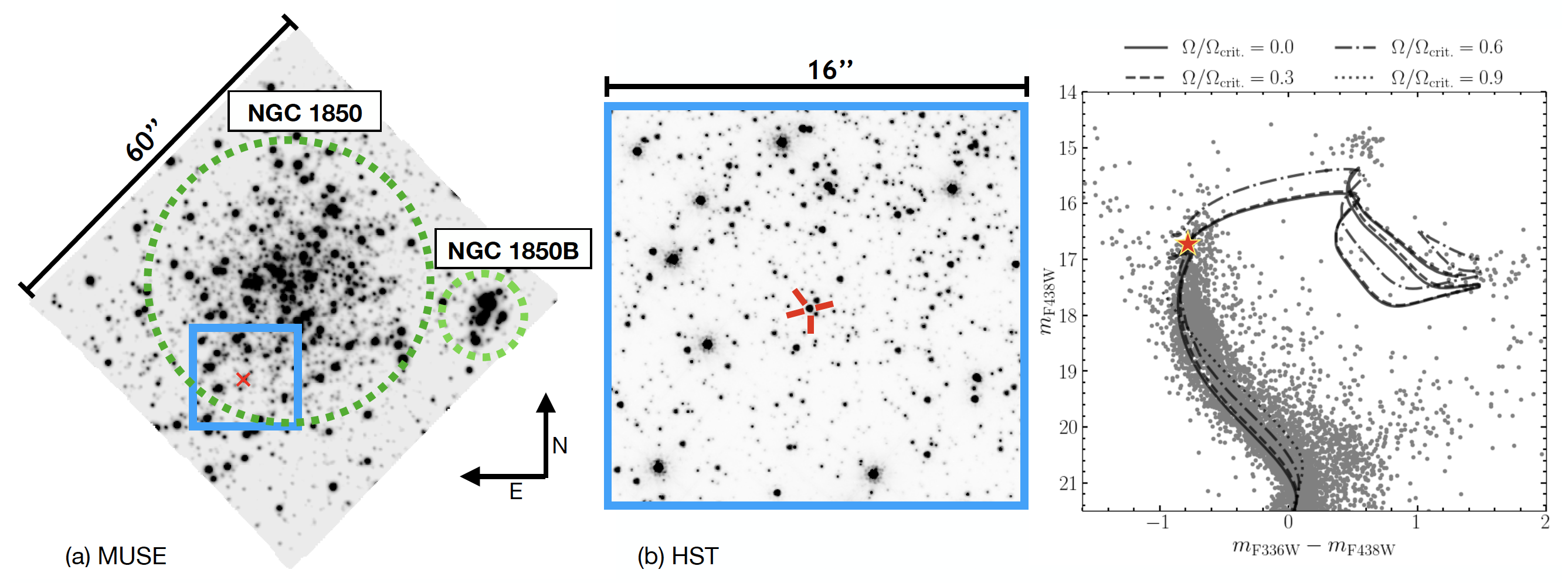}
    \caption{a) The central pointing of NGC~1850, as seen by MUSE. Green circles identify the position of NGC~1850 as well as the younger cluster NGC~1850B. The red cross indicates the target star. An inset of the region in the blue square is displayed in panel b). The image is taken from an archival HST/WFC3 image in the F814W band. The position of the target star is highlighted by red arrows. (c) Optical CMD of NGC 1850 from the archival HST data by \citet{Bastian2016}. The target star position is shown as an orange star. Solid, dashed and dotted lines represent a set of rotating and non-rotating MIST models.}
    \label{fig:fov}
\end{figure*}

The target star (RA: 77.1945$^{\circ}$ , Dec: $-68.7655^{\circ}$ [05:08:46.7 $-$68:45:55.6]) is located at a distance of about 18.8" (4.47 pc) from the center of NGC~1850 and within its effective radius $\bm{r}_\text{eff}$ = 20.5 $\pm$ 1.4" (4.97 $\pm$ 0.35 pc, \citealt{Correnti2017}). The position in the MUSE FOV is shown in Figure \ref{fig:fov}, panel a), while panel b) shows a 16" x 16" zoom of the region around the star from HST in the F814W band. The star is relatively bright, 16.7 mag in the F438W filter (16.6 mag in  F814W), and located on the main-sequence turn-off (MSTO) of NGC~1850 in the HST/WFC3 CMD (see Figure \ref{fig:fov}, right panel). We adopt a distance modulus of $(m-M)_{0}$ = 18.45 mag, an extinction E(B-V) = 0.1 and an age of the cluster of $\sim100$ Myr \citep{Bastian2016}. This allows us to determine the stellar mass of the visible star, as well as a first guess for the effective temperature $\bm{T}_\text{eff}$, surface gravity log(g) and metallicity, from the comparison with up-to-date theoretical models\footnote{The models adopted are for the evolution of a single star, however, it is likely that our target star had some form of interaction with its companion in the past. This may have an impact on the inferred mass, but we have not accounted for this effect, as binary evolution models are highly uncertain.}. Since stars in clusters of this age show significant rotation, we have used a set of rotating and non-rotating MIST models (Age=100 Myr and Fe/H=-0.2, \citealt{Gossage2019,mist2016}), deriving a mass for the visible component of $\sim4.9~M_{\odot}$. Figure \ref{fig:fov}, right panel, shows isochrones with different rotation rates, and the difference in mass at the magnitude level of the target star does not exceed $\pm$ 0.25 $M_{\odot}$. To evaluate the effect introduced by errors on the adopted parameters, we simulated a synthetic population of $2 \times 10^5$ stars, assuming gaussian distributions of (100 $\pm$ 20) Myr, (0.1 $\pm$ 0.03) mag, (18.45 $\pm$ 0.05) and (-0.20 $\pm$ 0.25), for age, extinction, distance modulus and metallicity, respectively. The mass distribution for stars at the MSTO level, according to the MIST isochrones, peaks at 4.9 $M_{\odot}$, with $\sigma$ = 0.3 $M_{\odot}$. Once photometric errors and different rotation rates are also taken into account, the uncertainty on the target star mass is $\sim$ $\pm$0.4 $M_{\odot}$.

For the target star we have a sample of 22 extracted spectra in total, with 17 having S/N greater than 10, while 5 showing a lower S/N (see Table \ref{tab:muse_data}). The 5 low S/N spectra come from the outer pointings, where this star is unfortunately located near one of the edges of the camera. In order to avoid introducing any spurious effects in the radial velocity data, we decided to discard these 5 spectra from the subsequent analysis. 

\begin{figure}
    \centering
	\includegraphics[width=0.475\textwidth]{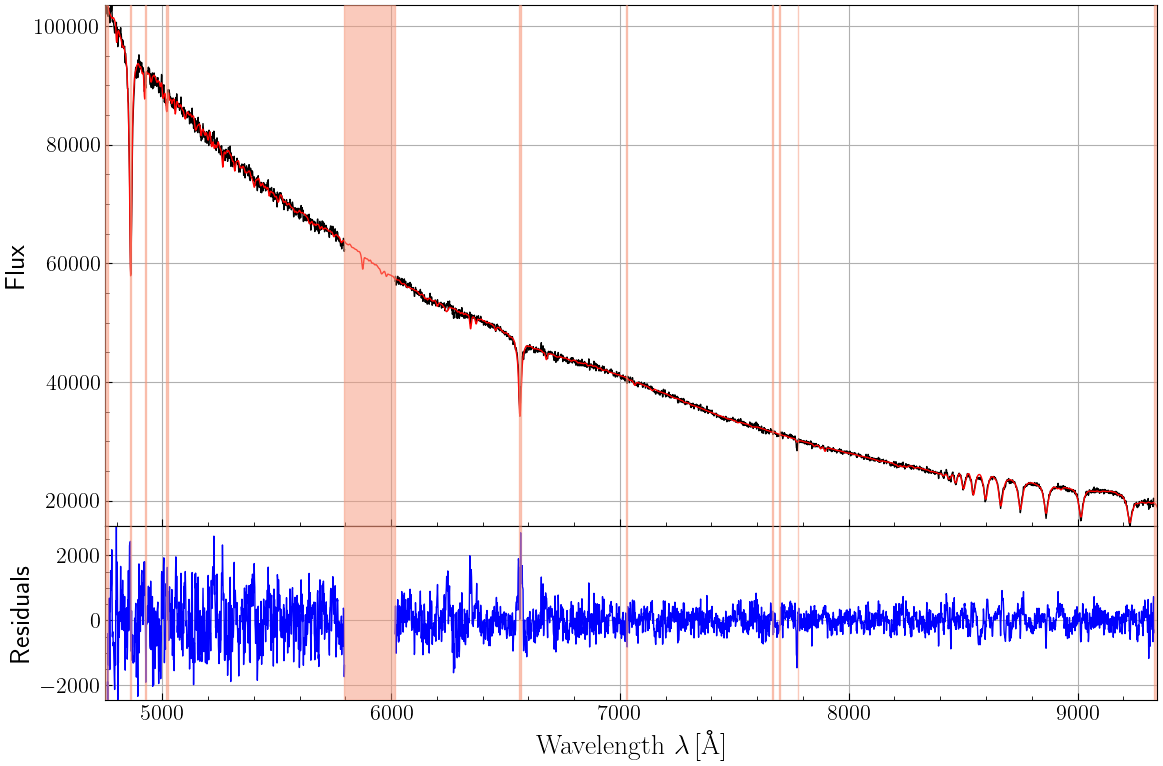}
    \caption{The combined, rest-framed, spectrum of our target star is shown in black. The best-fitting FERRE spectrum is over-plotted in red. The bottom panel shows the residuals after subtracting the best fit model to the data. Pink shaded regions are the spectral intervals masked out during the analysis due, for example, to the sodium emission of the laser in the AO mode, and the diffuse nebulosity associated with NGC~1850B.}
    \label{fig:spectrum}
\end{figure}

For analysing the spectra we made use of \textsc{Spexxy} \citep{Husser2016}, a software that determines radial velocities as well as stellar parameters (effective temperature, metallicity) using full-spectrum fitting against a set of templates. For the synthetic templates we adopted the \textsc{Ferre} library \citep{allendeprieto2018}. This library contains model stellar spectra for B-type stars, which are necessary when dealing with the young stars in NGC~1850.
The FOV of NGC~1850 is contaminated by diffuse nebular emission, associated to the much younger ($\sim$5~Myr) cluster NGC~1850B (shown as a small green circle in panel a) of Figure \ref{fig:fov}), hence all the spectral ranges that could potentially be contaminated by such an emission were masked out in our spectral analysis. The combined, rest-framed, spectrum of our target star is presented in black in Figure \ref{fig:spectrum}, with the best-fit model derived with \textsc{Spexxy} overplotted in red. It looks like a standard B-type star spectrum but we find large changes in radial velocity of up to 307.9 km/s between epochs. The MUSE radial velocities and the corresponding S/N are listed in Table \ref{tab:muse_data}. We have also verified that the variation we see in radial velocities comes effectively from the source, by analyzing the only bright star ($\sim$0.5 mag fainter) in its proximity (RA: 77.1943$^{\circ}$, Dec: -68.7654$^{\circ}$, to the north-west with respect to the source in panel b) of Figure~\ref{fig:fov}) that could in principle contaminate its spectrum. This star, indeed, does not show any sign of variability{\footnote{It has a probability of 12\% to be variable, meaning that the radial velocity variations we observe for this star are not significant compared to the uncertainties of the individual RV measurements.}}, therefore confirming that our target star is genuinely variable.
By assuming a mean metallicity [Fe/H] = -0.20, consistent with previous estimates for the cluster iron content, and log(g)=3.57 as derived from the isochrones, we measured $\bm{T}_\text{eff}$ $\sim$ 14500 $\pm$ 500 K, which is again expected for a B-type star.

\begin{table}
\centering
\caption{MUSE radial velocity measurements for the target star and spectral S/N per pixel, averaged over the MUSE wavelength range. Radial velocities derived from spectra with S/N smaller than 10 are also included for completeness but they were excluded from the analysis to avoid introducing any spurious variability.}
\begin{tabular}{c c c c}
\hline
Time (MJD) & V$_{R}$ (km/s) & $\sigma$ V$_{R}$ (km/s) & S/N  \\
\hline
58550.02867354 &	131.6 &	8.7	& 37.5 \\
59201.25318966 &	151.9 &	8.9	& 36.1 \\
59175.16932389 &	136.7 &	9.1	& 35.2 \\
59203.14316930 &	401.1 &	9.0	& 34.4 \\
58497.08534751 &	360.7 &	8.7	& 33.9 \\
58556.01231551 &	135.0 &	10.4 &	33.8 \\
59176.13916114 &	148.2 &	10.8 &	32.1 \\
58498.15614836 &	364.4 &	10.7 &	31.6 \\
59190.19805799 &	143.0 &	9.6	& 31.6 \\
58553.01788807 &	399.5 &	9.8	& 28.7 \\
59174.27808058 &	248.7. & 10.5 &	28.6 \\
59251.14817479 &	139.6 &	11.8 & 27.5 \\
59175.29610617 &	96.3. &	13.4 &	25.0 \\
59176.30463698 &	178.9 &	12.6 &	24.7 \\
59174.32203504 &	250.9 &	21.3 &	14.9 \\
59177.30703879 &	340.7 &	21.9 &	13.8 \\
58556.02541888 &	93.2 &	19.6 &	12.2 \\
\hline
58550.04180262 &	147.7 &	20.2 &	9.1 \\
58498.17027805 &	402.5 &	21.6 &	8.4 \\
58553.03123202 &	435.9 &	37.3 &	4.8 \\
59176.15348631 &	93.3 &	60.2 &	2.1 \\
59175.17711612. &	173.4 &	62.1 &	1.8 \\
\hline
\end{tabular}
\label{tab:muse_data}
\end{table}

\section{Results}
\subsection{Radial Velocities}
\begin{figure}
    \centering
\includegraphics[width=0.475\textwidth]{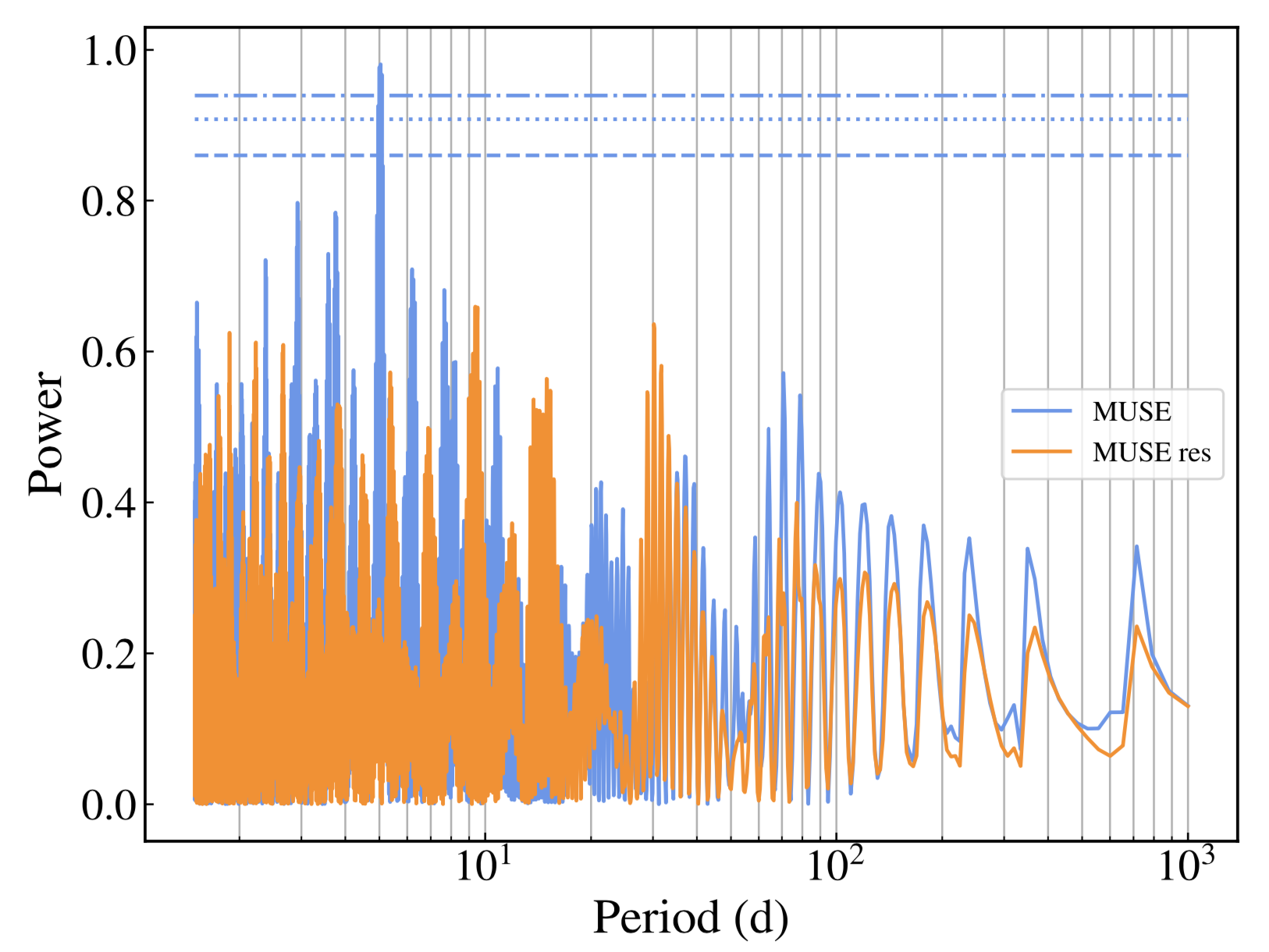}
    \caption{Generalized Lomb-Scargle periodogram of the MUSE radial velocities of the target star, before (blue curve) and after (orange curve) the subtraction of the main peak at $\sim$5.04 days. Horizontal lines (from bottom to top) indicate the power corresponding to a 1\%, 0.1\%, and 0.01\% false alarm probability. All peaks under these curves have a high probability of being unreal.}
    \label{fig:GLS}
\end{figure}

To constrain the orbital properties (the five standard Keplerian parameters plus the velocity semi-amplitude) of this binary system as well as to estimate the minimum mass of the unseen companion, we made use of THE JOKER \citep{joker2017,joker2020}. This software is a custom Monte Carlo sampler for sparse radial velocity measurements of two-body systems and can produce posterior samples for orbital parameters even when the likelihood function is poorly behaved. It is ideal in our case where only 17 radial velocity measurements, not equally sampled in time, are available. We generated $2^{29}$ prior samples\footnote{We refer the reader to the documentation for JokerPrior.default for more details on the prior distributions.} for the period range 0.3 d to 4096 d. We requested a maximum of 256 posterior samples, by assuming a gaussian distribution for the cluster systemic velocity with 250 km/s and 5 km/s as mean and dispersion, respectively \citep{Kamann2021}.
We found a unimodal solution with the binary system having a relatively short orbital period of P = 5.0402 days and moving almost on a circular orbit (very low eccentricity e). We ran THE JOKER again, this time limiting the possible solutions around the identified orbital period, specifically between 2 and 10 days, to increase the resolution of the grid. In addition, we also used the generalized Lomb-Scargle \citep[GLS,][]{zechmeister2009} periodogram to analyse the radial velocity curve, and we found that aliasing is not a problem in this case, as a clear periodicity can be isolated robustly from the data. This is shown as a blue curve in Figure \ref{fig:GLS}, where a definite peak can be identified, with power close to 1. The overplotted orange curve instead shows the periodogram after the subtraction of the main peak.

\begin{figure}
    \centering
\includegraphics[width=0.475\textwidth]{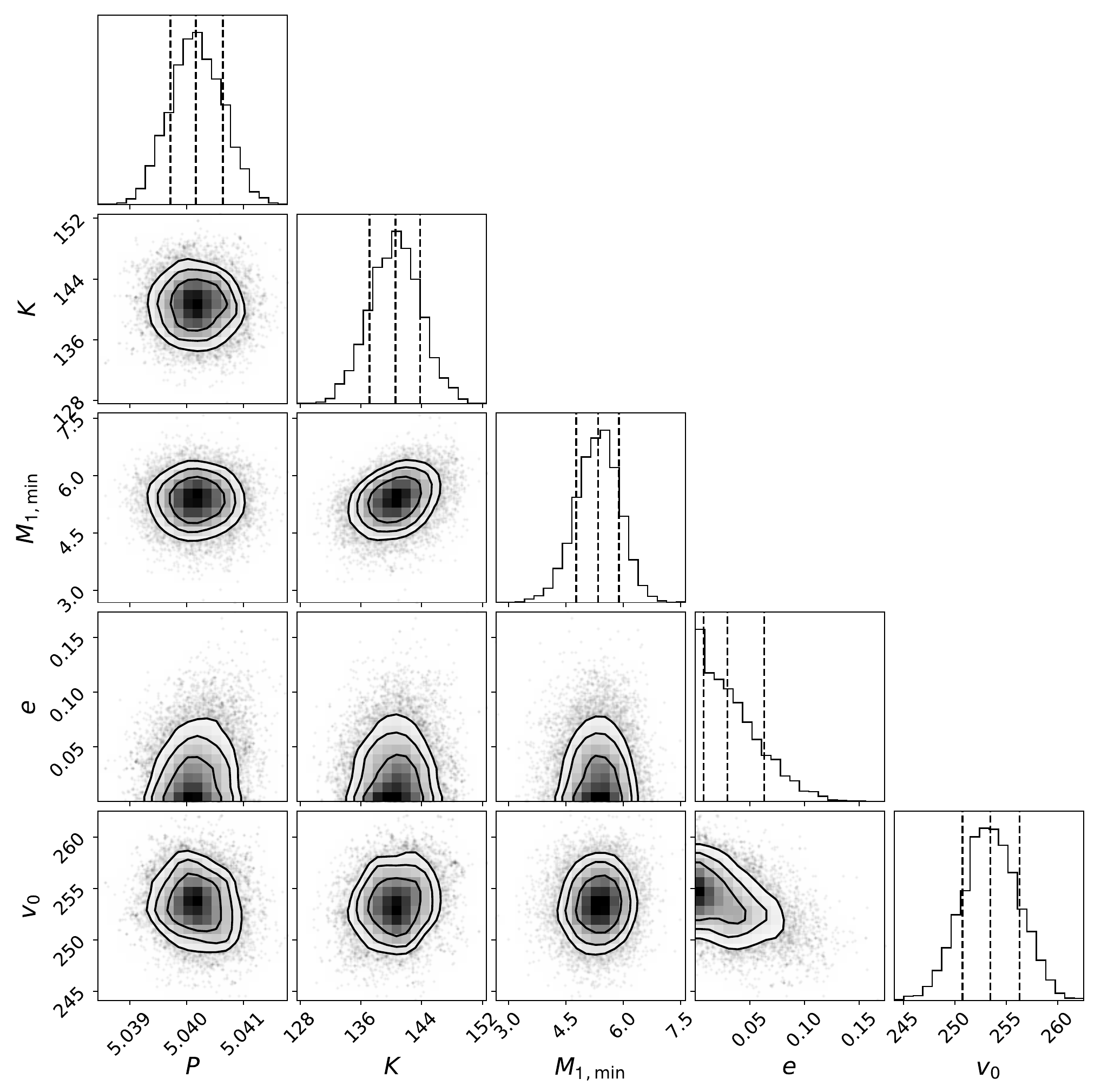}
    \caption{Corner plots showing the one- and two-dimensional projections of the posterior probability distributions for all the parameters of the binary system derived from THE JOKER+MCMC: Period P [d], velocity semi-amplitude K [km/s], minimum unseen companion mass $M_{1,{\rm min}}$ [${M_\odot}$], eccentricity e, and barycenter velocity $v_{0}$ [km/s]. The contours show the 1, 2 and 3$\sigma$ levels.}
    \label{fig:joker}
\end{figure}

To generate more posterior samples and reliably estimate uncertainties on the orbital parameters, as well as a minimum mass for the unseen companion (by assuming an inclination = 90$^{\circ}$ for an orbit seen edge-on), we performed a Monte Carlo Marchov Chain (\textsc{MCMC}) analysis within THE JOKER. The (marginalized) posterior means and 1$\sigma$ uncertainties are shown in Figure~\ref{fig:joker} as corner plots, as well as listed in Table~\ref{table:binary_results}. As can be seen, the predicted minimum mass of the unseen source (i.e., fainter star) is $\sim$ 5.34 $M_{\odot}$, already higher than the mass of the visible binary component (i.e., brighter star). The same lower limit on the mass of the unseen companion can be derived analytically from the observed velocity amplitude and period, by using the mass function in equations 1 and 2 of \citet{ducati2011}, and assuming a mass for the observed binary component. 

This result, corroborated by the visual inspection of both the CMD and the MUSE spectra, represents the first clear indication of this object being compact and dark rather than luminous, and in particular a stellar-mass BH, due to its estimated minimum mass higher than any possible neutron star ($M \sim 3 M_{\odot}$, \citealt{NS2001}). Furthermore, the fitted radial velocity of the binary barycenter ($v_{0}$ $\sim$ 253 km/s) is in good agreement with the systemic velocity of the cluster, making the target star as a high probability cluster member\footnote{Unfortunately, while our target star is included in Gaia EDR3 \citep{2021A&A...649A...1G}, its astrometric solution fails several of the quality criteria listed in \citet{Fabricius2021}, preventing us to reach any conclusion about the membership to the cluster from its proper motion.}. To verify that this is not a consequence of our prior (which assumed a Gaussian distribution of velocities around the cluster mean), we ran THE JOKER again by assuming a velocity distribution similar to that of the LMC field in the region, sensibly increasing the number of possible solutions for the binary. We found that, even in this case, the software converges to the same best-fit solution for the radial velocity barycenter, confirming this binary is most likely a cluster member.

\subsection{Light Curves}
Radial velocity measurements alone cannot provide us with any constraints on the inclination of the orbit of the binary system, hence only a minimum mass for the unseen companion can be estimated. Only in special configurations, i.e. eclipsing binaries, where photometric light curves are also available, the inclination of the orbit can be inferred in a reliable way.

Fortunately, this star is included in the catalog of LMC variable stars compiled by OGLE (3rd release) as OGLE-LMC-DPV-039 and classified as double periodic variable \citep{poleski2010}. It shows two periods, a shorter one at 5.040495 d and a $\sim$30 times longer one at $\sim$156 d. This class of objects is rather enigmatic, in fact, if the shortest period is associated to the orbital motion, the origin of the longer periodic variation is still unclear \citep{Mennickent2003}. 
The orbital period of our target star derived by THE JOKER ($5.0402^{+0.0004}_{-0.0004}$ d) using radial velocity variations perfectly matches the short period found photometrically by OGLE, confirming that we are indeed looking at the same star.
This star is also in the OGLE 4th release as OGLE-LMC-ECL-29851 and \citet{pawlak2016} additionally classified it as an eclipsing binary. By visually inspecting the light curves, we were unable to clearly see an eclipse in the data and by applying the GLS periodogram to the light curves, we identified a significant peak at about 2.52 days, corresponding to $P_{orb}/2$, with no peaks around the orbital period. This is an unambiguous property of an ellipsoidal variable, where the light curves show two maxima and minima per orbit - two cycles for every one cycle of the velocity curve. This finding could then suggest that the star was misclassified as an eclipsing binary, rather than an ellipsoidal variable, in the OGLE IV catalog.

By using the latest version of \textsc{PHOEBE} (v2.3) \citep{conroy2020}, a software for modelling the light and radial velocity curves of binaries, we were able to put further constraints on the system and also confirm previous findings. Namely: 1) this system is, in fact, not an eclipsing binary as no eclipses are present in the light curves. Given how symmetrical the light curves are, if the system was an eclipsing binary, the two stars would have to have similar radii and luminosities. However, the star we see is clearly a single lined binary (SB1) so it must be significantly brighter than its unseen companion. 2) it is most likely in a semi-detached binary system with the visible star filling its Roche Lobe. In this configuration the B-type star gets distorted by the tidal influence of its orbiting companion and it takes on an elongated or ellipsoidal shape, becoming an ellipsoidal variable. The light variability we see in the OGLE light curves is caused mainly by the change in the apparent surface area as the star orbits around its companion. This peculiar feature has been extensively used in the literature, mainly for X-ray binaries and cataclysmic variables (CVs), as it allows to constrain the inclination of the systems and the masses of the compact objects (see \citealt{Orosz2003} and \citealt{Avni1975} for more details). 3) an orbital inclination larger than 50$^{\circ}$ can be excluded with a high confidence level, either in the case of a luminous or a dark companion, because the amplitude of the optical variations and the different depth between primary and secondary minima would not match the observational data (unless properties inconsistent with its position in the CMD are assumed for the B-type star). More details can be found in Appendix \ref{app:tests_1}. Given the configuration of the system, we will refer, hereafter, to the massive (probably dark) object as primary component of the binary, while to the B-type star which has started to fill its Roche Lobe as secondary component. 

An orbital inclination of i=50\degr would imply a true mass for the companion higher than the minimum mass of $5.34~M_{\odot}$ estimated by \textsc{THE JOKER}. This can be easily derived using Eq. 1 and 2 in \citet{ducati2011}, hence the mass ratio q = $M_{2}/M_{1}$ of the binary would be much smaller than 1. This finding unequivocally rules out the possibility that the primary star could be luminous. If it is the case, indeed, it would be significantly brighter that the secondary, thus making it visible in the HST photometry, as well as in the MUSE spectra. Hence, we can firmly state that the primary star is a dark compact object, specifically a BH, the first directly dynamically detected in a young massive cluster to date, and we call it NGC 1850 BH1. 

\begin{figure}
    \centering
	\includegraphics[width=0.475\textwidth]{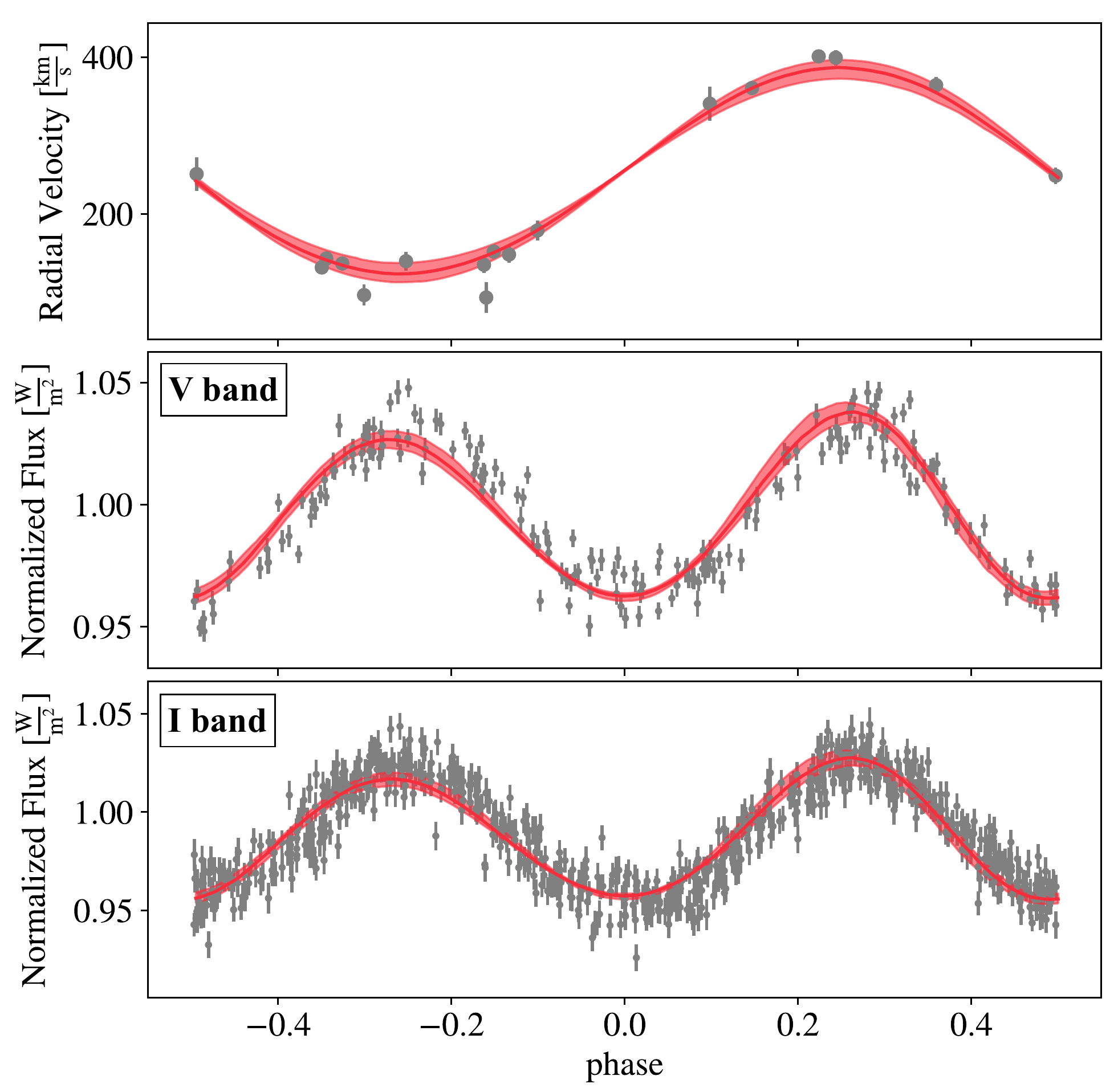}
    \caption{MUSE radial velocity curve (top) and OGLE light curves (centre: V-band, bottom: I-band) of the target star, phase-folded using a period of $P=5.04~{\rm d}$ (cf. Table~\ref{table:binary_results}), are shown as grey dotes. The red solid line represents the best-fit model from \textsc{PHOEBE}. The red shaded areas indicate the 1$\sigma$ uncertainties of the adopted parameters. The light curves show two maxima and minima per orbit - two cycles for every one cycle of the velocity curve - an unambiguous property of an ellipsoidal variable.}
    \label{fig:curves}
\end{figure} 

\begin{figure*}
    \centering
	\includegraphics[width=0.8\textwidth]{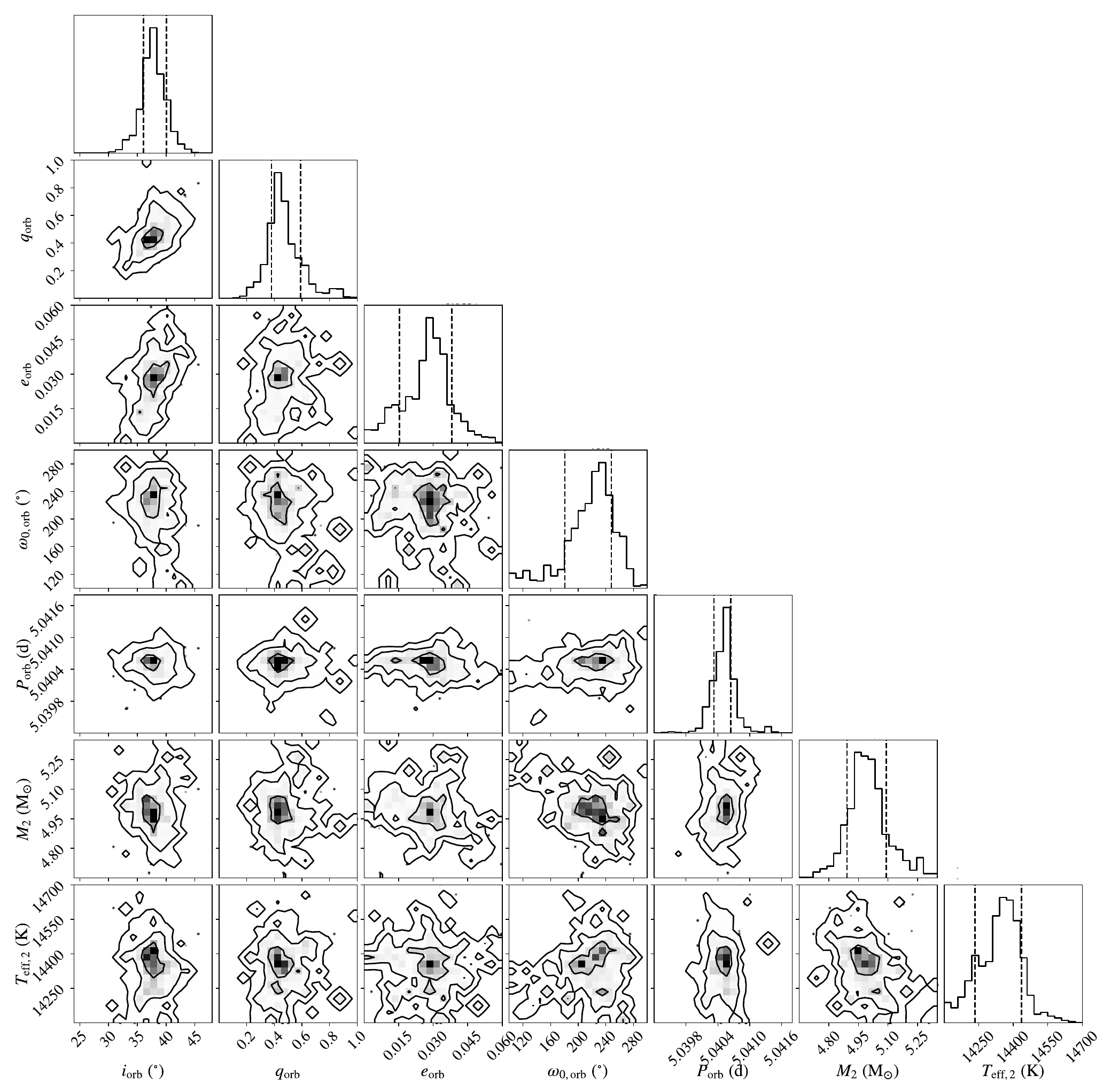}
    \caption{Corner plots showing the one- and two-dimensional projections of the posterior probability distributions for all the parameters of the binary system derived from \textsc{PHOEBE}+MCMC: inclination i [$^{\circ}$], mass ratio q, eccentricity e, argument of periastron $\omega_{0}$ [$^{\circ}$], Period P [d], Mass [$M_{\odot}$] and $\bm{T}_\text{eff}$ [K] of the visible component. The contours correspond to the 1, 2 and 3$\sigma$ levels.}
    \label{fig:phoebe}
\end{figure*}

In Table \ref{table:binary_results} we present the main properties of the binary system we derived from \textsc{PHOEBE}. For the modelling of the radial velocity and light curves we assigned the secondary star of the system the properties inferred from our analysis of the MUSE spectra, i.e. a MSTO star with $\bm{T}_\text{eff}$=14500 K and mass of 4.9 M${_\odot}$. We used the option "distortion method=none" which only accounts for the gravitational influence of the compact source (the BH companion) and assume it is otherwise completely dark and transparent.\footnote{To simulate the presence of a BH we alternatively set the lowest temperature allowed by the code ($\bm{T}_\text{eff}$=300 K) and a radius R = $3\times10^{-6}$~$R_{\odot}$, as done in \citet{Jayasinghe2021}, finding consistent results.} In particular, we performed an MCMC run (nwalkers=48, niters=1000, burnin=226) with \textsc{emcee} \citep{emcee2019} by fitting over the following parameters: inclination and period of the binary, mass ratio, eccentricity, argument of periastron and the mass and temperature of the B-type star. For those parameters that have been partially constrained by the previous analyzes, we have imposed a Gaussian distribution around their values while for the others (i.e. orbital inclination and mass ratio) for which we do not have any constraints we have assumed a uniform prior over the entire allowed range. The OGLE V and I band light curves, as well as the MUSE radial velocities are presented as grey dots in Figure~\ref{fig:curves}, with the best-fit model from \textsc{PHOEBE} overplotted as a solid red line. The red shaded areas represent the 1$\sigma$ uncertainties from the MCMC run propagated toward the best-fit model. The results (i = $37.9^{+2.2}_{-1.9}$$^{\circ}$ and q = $M_{1}/M_{2}$ = $0.45^{+0.14}_{-0.07}$) are also shown as corner plots in Figure \ref{fig:phoebe}, confirming the detection of a $11.1_{-2.4}^{+2.1}$ $M_{\odot}$ BH orbiting a MSTO star in NGC 1850. If the inclination of the system is indeed confirmed by further studies, this represents one of the few massive BHs detected so far and its relative mass uncertainty of 20\% (see Table \ref{table:binary_results}) makes it one of the most accurate measurements available in the literature \citep{Miller-Jones2021}.

In Appendix \ref{app:tests_2} we test the hypothesis that the OGLE light curve of our target star might be contaminated by the flux of the bright source in its proximity (see Figure \ref{fig:fov}, panel b). We apply PHOEBE to the new light curves to understand how strongly this may affect the main physical properties of the binary system we have derived in this Section. In summary, we obtain a slightly higher inclination and reduced mass of the BH of $9.2^{+1.6}_{-2.2}$ $M_{\odot}$, still within the uncertainties of the mass estimate given in Table \ref{table:binary_results}. 

\begin{table}
 \caption{Binary system properties, from \textsc{THE JOKER} and \textsc{PHOEBE}.}
 \label{table:binary_results}
 \centering
 \begin{tabular}{r l}
 \hline\rule{0pt}{2.3ex}
 THE JOKER + MCMC\\[2pt]
 Period $P$ & $5.0402 \pm 0.0004$ d \\
 Velocity semi-amplitude $K$ & $140.40^{+3.31}_{-3.42}$ km/s\\
 Barycentric radial velocity $v_0$ & $253.30^{+2.59}_{-2.44}$ km/s \\
 Minimum companion mass $M_{1} \sin(i)$ & $5.34^{+0.55}_{-0.59}$ $M_{\odot}$\\
 \hline\rule{0pt}{2.1ex}
 PHOEBE + MCMC\\[2pt]
 Inclination $i$ & $37.9^{+2.2}_{-1.9}$ degrees\\
 Mass ratio $q$ & $0.45^{+0.14}_{-0.07}$\\
 Eccentricity $e$ & $0.029^{+0.010}_{-0.014}$ \\
 Argument of periastron $\omega_{0}$ & $222^{+27}_{-41}$ degrees\\
 Secondary mass $M_{2}$ & $4.98^{+0.10}_{-0.10}$ $M_{\odot}$\\
 Effective Temperature $\bm{T}_\text{eff}$ & $14353^{+83}_{-119}$ K\\
 Semi-major axis $a$ & $31.2^{+1.3}_{-1.6}$ $R_{\odot}$\\
 Companion mass $M_{1}$ & $11.1_{-2.4}^{+2.1}$ $M_{\odot}$\\
 \hline
 \end{tabular}
\end{table}

\section{Searching for an X-ray counterpart}
Besides the strong evidence supporting the detection of a BH in NGC~1850 coming from the modelling of the radial velocity and light curves, we have searched for independent probes to support our interpretation.
When a BH is in a binary system, intense X-ray emission can arise from the accretion disk and the corona \citep[e.g,][]{1995xrbi.nasa..126T}. The analysis of the X-ray emission from compact objects accreting material from their companions can thus shed important light on the accretion process, the outflow, and all the physical processes occurring in the system. In such a case, the emission can provide us with further evidence for the presence of a BH. Following the review by \citet{2006ARA&A..44...49R}, the typically observed X-ray luminosity of BHs in binary systems in their quiescent state ranges from 10$^{30.5}\,$erg/sec to 10$^{33.5}\,$erg/sec, and their spectra are typically dominated by a power law component with $\Gamma$ between 1.5 and 2.1. Unfortunately, NGC 1850 is not the ideal target for X-ray studies as the cluster is rather compact and distant. Also, the region where the BH lies in the cluster is very close to the center, where crowding represents a critical issue.
Adopting a distance to NGC~1850 of 48$\,$kpc, and using PIMMS v. 4.11\footnote{https://cxc.harvard.edu/toolkit/pimms.jsp}, the nominal range of X-ray luminosity and spectral index result in an expected count rate ranging from $6.2\;x\;10^{-7}$ counts/sec to $5\;x\;10^{-4}$ counts/sec. Except the faintest limit, these count-rates result in a detectable, even if faint, source in X-ray observations with an effective exposure of a few hundreds ksec.
    
Among the available X-ray telescopes, only Chandra, with 13 ACIS-S observations within 10$\,$arcmin from the location of the target star, for a total exposure time of 414$\,$ksec (P.I. Williams and Portegies Zwart, the latter only for Obs. ID 3810, \citealt{2018ApJ...865L..13W}), could be able to detect such a kind of emission. Table \ref{tab:chandraobs} shows the Obs.ID, the exposures, and the aim points of the selected observations. 

\begin{table}
	\centering
	\caption{Chandra observations of NGC~1850 analyzed in this work}
	\label{tab:chandraobs}
	\begin{tabular}{cccc} 
		\hline
		Obs.ID & Exposure & RA & Dec \\
		       & ksec     & J2000   & J2000     \\		
        \hline
		3810 &    29.67&  05:08:44 & -68:45:36 \\
		18018&    39.54&  05:08:59 & -68:43:34 \\ 
		18019&    59.28&  05:08:59 & -68:43:34 \\
		18020&    27.19&  05:08:59 & -68:43:34 \\
		19921&    16.85&  05:08:59 & -68:43:34 \\
		19922&    41.43&  05:08:59 & -68:43:34 \\
		19923&    58.30&  05:08:59 & -68:43:34 \\
		20042&    19.80&  05:08:59 & -68:43:34 \\
		20053&    11.20&  05:08:59 & -68:43:34 \\
		20058&    43.79&  05:08:59 & -68:43:34 \\
		20067&    29.68&  05:08:59 & -68:43:34 \\
		20074&    31.17&  05:08:59 & -68:43:34 \\
		\hline
	\end{tabular}
\end{table}

\begin{figure*}
	\includegraphics[width=7cm]{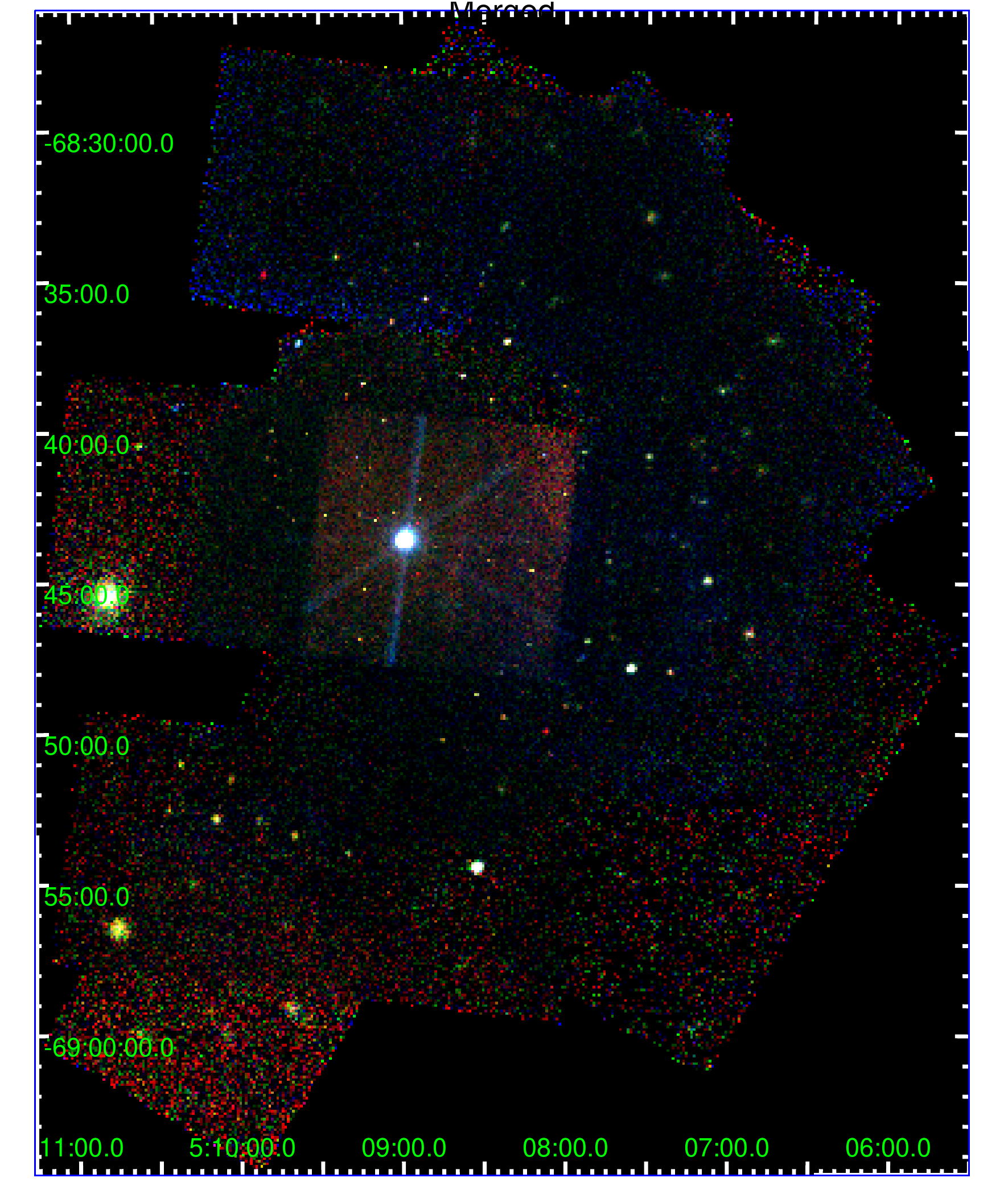}
    \includegraphics[width=7.28cm]{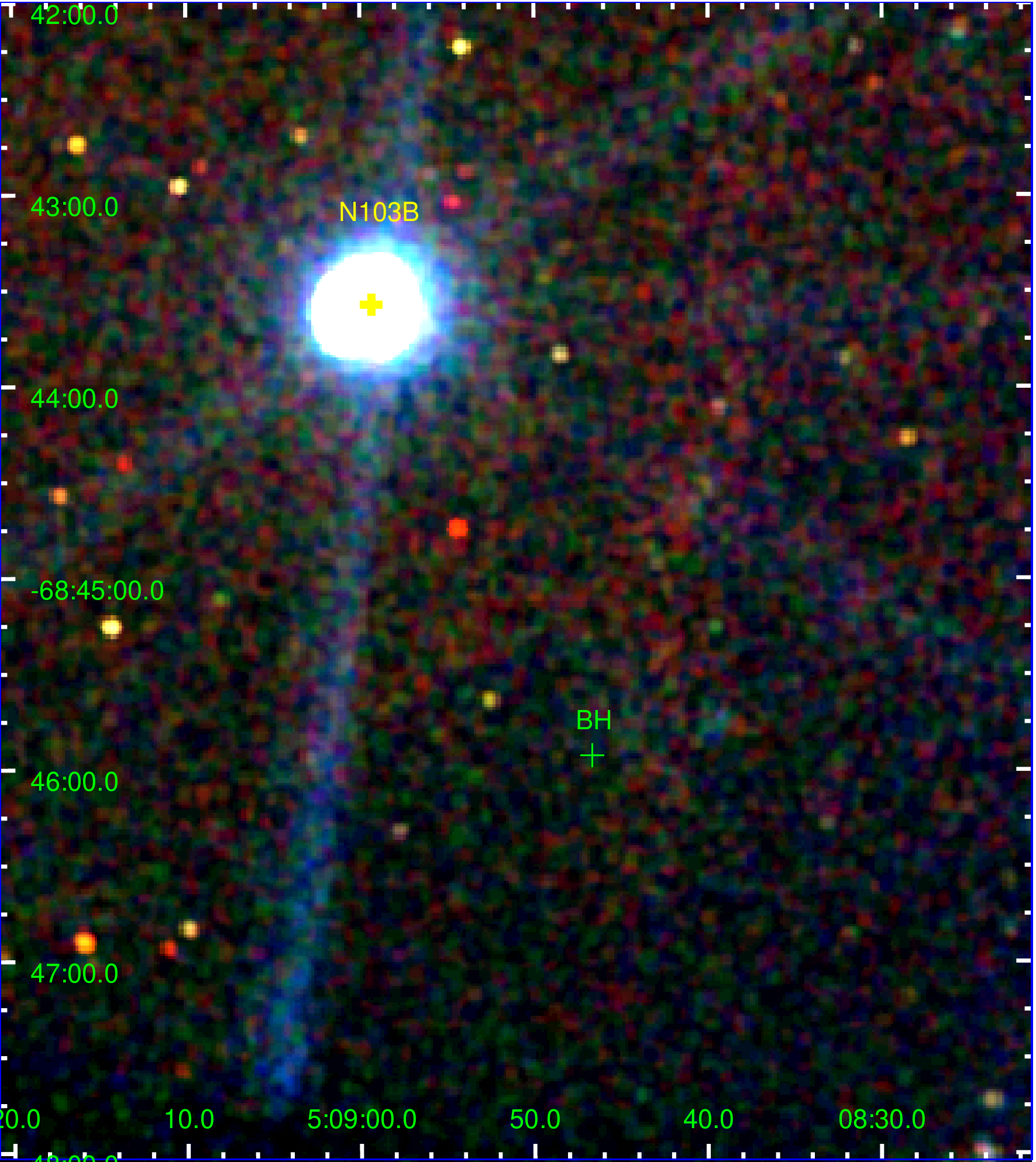}
    \caption{Merged RGB Chandra/ACIS-S images of NGC~1850 analyzed in this work, with events in the hard bands in red, medium band in green, and soft band in blue. The left panel shows the whole field, while the right panel is centered on the position of the candidate black hole and the bright SNR N103B.}
    \label{fig:acisfields}
\end{figure*}

We first reprocessed the available primary data distribution files to produce new Level 2 event files using the CIAO script {\it chandra\_repro} \citep{Fruscione2006SPIE.6270E.1VF}. We then followed the procedure described in the ``Correcting Absolute Astrometry'' CIAO thread to correct the astrometry of each observation by aligning it on the deepest ACIS-S observation. The observations were thus merged using the {\it merge\_obs} tool. Exposure maps in the broad (0.5-7.0$\,$keV), soft (0.5-1.2$\,$keV), medium (1.2-2.0$\,$keV), and hard (2.0-7.0$\,$keV) bands were calculated using the standard CIAO tools {\it asphist}, {\it mkinstmap}, and {\it mkexpmap}. Fig. \ref{fig:acisfields} shows a RGB image of the whole field and a zoom into the region where the target star, hence the candidate BH, and the supernova remnant (SNR) N103B are located.

A detailed detection and characterization of all the X-ray sources observed in the combined image is beyond the scope of this work. The strategy we adopted was, in fact, only aimed at investigating the presence (or absence) of a significant X-ray source at the position of the binary system and deriving its count-rate. We first performed the source detection in the four bands using the CIAO tool {\it wavdetect}, with an adopted detection threshold of $10^{-4}$. This choice resulted in 494 sources detected in the broad band, 448 in the soft band, 467 in the medium band, and 339 in the hard band. After a visual inspection of the detections, a list of 1105 unique candidate sources was compiled. We then used the IDL software ACIS Extract \citep[AE,][]{BroosTFG2010} with the aim of validating these sources. Since no source has been found by {\it wavdetect} at the position of the BH, we inserted into the catalogue by hand a source at its position in order to let AE check the presence of any valid source at that position.

AE defines an individual photon extraction region for each source calculating the PSF at 1.5$\,$keV, accounting for crowding by reducing the extraction region for crowded sources. The individual background is estimated by using an annular region centered on each source, with the inner radius equal to 1.1 the 99\% of the PSF, and the outer radius large enough to encompass 100 background photons. For crowded regions, AE calculates the background from a model that accounts for the presence of nearby sources. In order to improve this model, the background must be estimated with several iterations. We also used AE to correct source positions, adopting for on-axis sources ($\Theta \leq 5^{\prime}$) the {\it mean data position}, which is obtained from the centroid of the extracted events, the {\it correlation position} for the off-axis sources, which is obtained by interpolating the PSF with the events distribution, and the {\it maximum likelihood position} for the crowded sources, which is calculated from the maximum likelihood image of source neighborhood.
    
Once photons extraction and the estimate of the background was repeated after sources relocation, we calculated the parameter {\it prob\_no\_source} ($P_B$), which provides the probability of the source being a background fluctuation and thus an estimate of its reliability (reliable sources typically have $P_B < 0.01$). At the position of the binary system, AE found a marginally reliable source in the soft band ($P_B=0.003$\footnote{which roughly corresponds to the 35\% quantile of the $P_B$ values in the soft band}), which lies 0.45$\,$arcsec to the east of the nominal position of the putative BH. By checking the HST image, no sources other than our binary system may be responsible for such an emission. Furthermore, the separation we observe between the optical and X-ray counterpart is compatible with the astrometric offset between the two catalogs.\footnote{The mean astrometric offset between the two catalogs was measured using only the brightest sources. We identified 54 closest coincidences, adopting a tolerance radius of 10".}
Despite its reliability, only 2.7 net counts are detected in the soft band (with a mean photon energy of 0.64$\,$keV), which however corresponds to a count-rate of $1.87 \times 10^{-5}$ in the soft band, and thus, using PIMMS, to an X-ray luminosity of about $10^{33}$\, erg/sec.
Thus the result from the X-ray analysis would be consistent with a BH in a quiescent state. We emphasize, however, that a further characterization of the source is not possible, given its very few net counts.

\section{Discussion and Conclusions}
In this paper we exploit MUSE radial velocities and OGLE-IV light curves to report the detection of a short period (P = 5.04 d) binary system in NGC 1850, made-up by a $\sim\;$4.9~$M_{\odot}$ MSTO star and a $\sim$11~$M_{\odot}$ BH (NGC 1850 BH1), under the assumption of a semi-detached configuration. 
The membership to NGC 1850 was established by analyzing the barycentric radial velocity, the distance to the cluster centre, and the properties of the visible star (i.e. sharing the same turn-off mass and [Fe/H] as NGC 1850). The analysis of Chandra X-ray data revealed a faint but marginally reliable source at the location of the binary system, which would be consistent with the presence of a BH in a quiescent state. 

Future studies of the cluster dynamics as well as the secondary stars' chemistry will help shedding light on the origin (primordial vs dynamically formed binary) of this intriguing system.
The fate of the binary is however quite uncertain. Based on the evidence gathered so far, we speculate that the system will likely experience a Roche-Lobe overflow, as soon as the B-star evolves out of the main-sequence. There will be a stable mass transfer and significant X-ray emission, generally leading to a widening of the binary. Mass transfer will likely end when most of the H envelope of the donor star has either been transferred to the companion or been lost from the system, leaving an He star core \citep{1997A&A...327..620S,2008ASPC..391..323P}. If so, it will likely experience another phase of mass transfer (and X-ray emission) when the star burns He in shell, ending up as a BH + white dwarf system.

The detection of such a system has an important impact in different fields:
\\
i) \emph{GC studies}: Finding a BH in a stellar cluster of just $\sim$100 Myr represents the starting point in the construction of the BH initial mass function. Furthermore, it supports the search for the complete (dynamically detectable) population of BHs, allowing strict constraints to be placed on the BH retention fraction, a major uncertainty in GC models.\\
ii) \emph{The studies of binaries}: The physical properties of the visible star are very well constrained (e.g. mass, metallicity, age/evolutionary stage) by being member of a massive cluster. This precision can help understand the physics responsible for the long-term periodic variations observed in the light curves of double periodic variables across the Milky Way and the Magellanic Clouds.\\
iii) \emph{The search for non (or weakly) interacting compact sources:} Our finding validates both the power and reliability of the radial velocity approach as an important tool for detecting these types of systems in different environments, especially in GCs. This result represents a proof of concept and will be extended to a diverse sample of GCs (at a variety of ages) in the future.\\
iv) \emph{BH mass measurement}: Very few mass measurements are available in the literature for NSs and BHs. They are mainly biased toward accreting binary systems selected via radio, X-ray, and $\gamma$-ray data (see, for e.g., \citealt{Champion2008,Liu2006,Ozel2010,Farr2011}), and from the LIGO/Virgo detections of merging systems (see, for e.g., \citealt{Abbott2016a,Abbott2017}). Deriving the masses of non-interacting compact objects is a rare event (a few examples can be found in \citealt{Thompson2019} and \citealt{Jayasinghe2021}) but it really becomes a novelty when they reside in GCs. This represents the very first step towards an unbiased characterization of the BH mass distribution in clusters.

Finally, although the radial velocity method presented here is not sensitive to binaries made-up by two non-luminous components (like BH-NS or BH-BH binaries), which are considered the main GWs emitters, similar studies are crucial for: {\it i)} understanding all the evolutionary phases in between a massive binary star and a binary BH; {\it ii)} unveiling the population of BHs in massive clusters, as this is the most likely place in the Universe, due to the frequent dynamical encounters, where BH merger cascades could be possibly triggered.

\section*{Acknowledgements}
We thank the referee, Dr. Paul J. Callanan, for the careful reading of the paper. Insightful comments and suggestions helped us improve the manuscript.
The authors warmly thank Selma de Mink, Peter Jonker and Daniel Mata-Sanchez for their valuable comments on the draft. SS thanks A. Price-Whelan and K. Conroy for very helpful inputs and discussions on THE JOKER and PHOEBE, respectively.
SS, NB and ICZ acknowledge financial support from the European Research Council (ERC-CoG-646928, Multi-Pop). SK acknowledges funding from UKRI in the form of a Future Leaders Fellowship (grant no. MR/T022868/1). CU acknowledges the support of the Swedish Research Council, Vetenskapsr{\aa}det. MG acknowledges support from the Ministry of Science and Innovation through a Europa Excelencia grant (EUR2020-112157). VHB acknowledges the support of the Natural Sciences and Engineering Research Council of Canada (NSERC) through grant RGPIN-2020-05990. Based on observations made with ESO Telescopes at La Silla Paranal Observatory. Based on observations of the NASA/ESA Hubble Space Telescope, obtained from the data archive at the Space Telescope Science Institute. STScI is operated by the Association of Universities for Research in Astronomy, Inc. under NASA contract NAS5-26555.

\section*{Data Availability}
The MUSE data underlying the paper will be shared on reasonable request to the authors, while the Chandra data are available for download in the corresponding archive. The light curves are available on the OGLE website.

\bibliographystyle{mnras}
\bibliography{star224} 


\appendix
\section{Binary system configuration}
\label{app:tests_1}
The results reported in the main text for the inclination and the true mass of NGC 1850 BH1 are based on the assumption that the binary system is in a semi-detached configuration. This assumption is strongly supported by the shape of the light curves themselves, as well as by the properties of the visible star derived from the comparison with up-to-date stellar evolutionary models. For completeness, however, we mention here that a best-fit solution comparable (at least qualitatively) to the one presented in Figure \ref{fig:phoebe} can be obtained also under the assumption of a detached system, while assuming a lower value for log(g) (by $\sim$0.3 dex) compared to the one derived from our analysis. This value falls well outside the estimated uncertainties for the surface gravity of the star (if real, indeed, at fixed mass and $\bm{T}_\text{eff}$, the star would appear much brighter than actually observed, see the right panel of Figure \ref{fig:fov}). For these reasons we consider the detached configuration unlikely, hence we adopted the semi-detached one to model the light curves.

Nevertheless, this test has been extremely informative as it allowed us to get two important confirmations: 1) even assuming a detached system, the inclination of the binary does not decrease by much ($\sim$10\degr - 15\degr), hence the companion still needs to be massive ($> 8$~M$_{\odot}$), and no other possibility other than a BH can be considered. 2) an important source of uncertainty in modelling the light curves is the value adopted for log(g), which unfortunately we cannot measure directly from our MUSE spectra due to their low resolution. This modelling, indeed, would significantly benefit from this information coming from high resolution data of the source, which are currently unavailable. These observations will help in discriminating between the two aforementioned configurations, hence significantly reducing the uncertainties on both the inclination and the mass of the BH.

\section{OGLE photometry and its resolving power}
\label{app:tests_2}
OGLE is one of the largest sky variability surveys but it has a lower spatial resolution compared to HST, hence the photometric accuracy is limited in dense environments like the innermost regions of star clusters. For this reason, there is a concrete probability that the OGLE photometry cannot resolve sources which are relatively close to each other ($\sim$ 0.5 arcsec) on the sky. This could be the case with our target star, which is located within the effective radius of NGC 1850 and has a relatively bright star in its proximity (see Figure \ref{fig:fov}, panel b). The offset of $\sim$ 0.4 mag measured in the I band between OGLE and HST, in fact, might be an indication of such a contamination but to definitively answer this question we would need a light curve obtained from high spatial resolution photometry such as HST, which is currently unavailable.

\begin{figure}
    \centering
	\includegraphics[width=0.475\textwidth]{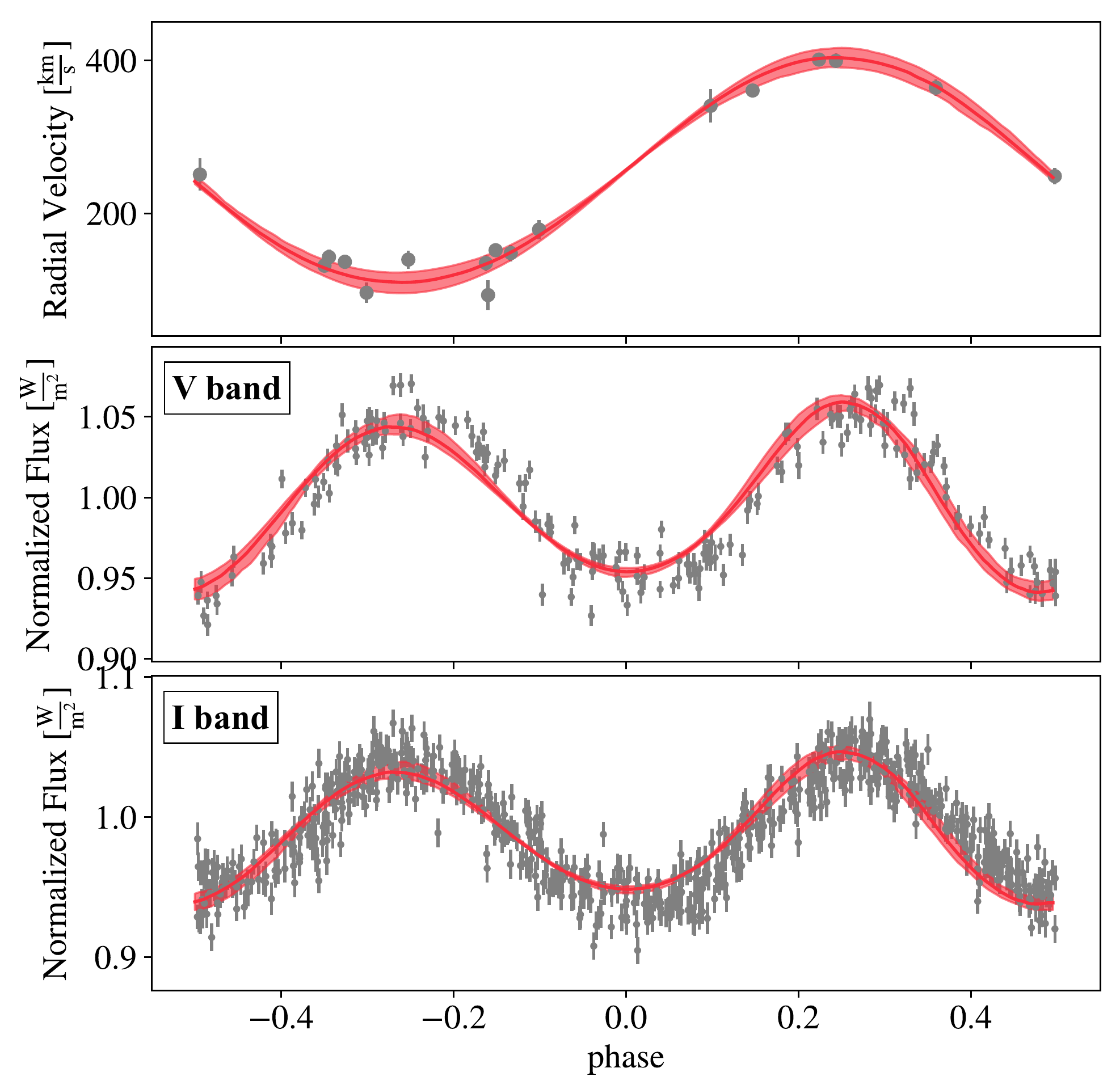}
    \caption{As in Figure \ref{fig:curves} but with the OGLE light curves decontaminated by the flux of the bright nearby star.}
    \label{fig:curves_deblended}
\end{figure}

To investigate the impact of a potential blend in the OGLE data on our results, we assumed the worst case, namely that OGLE was unable to resolve the two stars. We then subtracted the expected light contribution of the nearby star from every data point in the OGLE light curves. We used the HST zero points \citep{Sahu2014} to convert the WFC3 F814W magnitude of the nearby star to the Johnson-Cousins I band\footnote{The OGLE I and V bands are similar to the standard Johnson-Cousins I and V bands \citep{Udalski2015}}. The F555W band magnitude, which is not available in the HST catalog, was instead deduced from the comparison with a MIST isochrone.
After subtracting its flux from all measurements, we have derived new light curves, which show increased photometric modulation compared to the original ones (by $\sim$ 0.08-0.09 mag, see Figure \ref{fig:curves_deblended}). Interestingly, after the subtraction, the aforementioned offset between OGLE and HST cancelled out almost perfectly.

As a further test, we recovered the V and I band magnitudes from all extracted spectra and created photometric light curves from the MUSE data, using one isolated nearby star from OGLE as calibrator. Again the offset we observe relative to the magnitudes in the OGLE catalog is consistent with what one would expect if the two stars are blended in the latter.

To derive the main physical properties of the binary and see how they change when the original vs the new light curves are adopted, we run PHOEBE again, as done in Sec. 4.
The results are shown in Figure \ref{fig:curves_deblended} and Table \ref{table:binary_results_deblended} and can be directly compared with those in Figure \ref{fig:curves} and Table \ref{table:binary_results}, respectively. As can be seen, assuming that the new light curves are the most reliable, then the inclination of the binary system would increase by $\sim$12\degr and the mass ratio q would be of 0.53, leading to a BH mass of $9.2^{+1.6}_{-2.2}$ $M_{\odot}$, smaller than the previous estimate but still compatible within their uncertainties.

\begin{table}
 \caption{Binary system properties from \textsc{PHOEBE} when decontaminated light curves are considered.}
 \label{table:binary_results_deblended}
 \centering
 \begin{tabular}{r l}
 \hline\rule{0pt}{2.3ex}
 PHOEBE + MCMC\\[2pt]
  \hline\rule{0pt}{2.1ex}
 Period $P$ & $5.0405 \pm 0.0003$ d \\
 Inclination $i$ & $49.6^{+4.5}_{-2.0}$ degrees\\
 Mass ratio $q$ & $0.53^{+0.16}_{-0.07}$\\
 Eccentricity $e$ & $0.031^{+0.008}_{-0.007}$ \\
 Argument of periastron $\omega_{0}$ & $209^{+35}_{-31}$ degrees\\
 Secondary mass $M_{2}$ & $4.92^{+0.11}_{-0.15}$ $M_{\odot}$\\
 Effective Temperature $\bm{T}_\text{eff}$ & $14470^{+130}_{-120}$ K\\
 Semi-major axis $a$ & $29.8^{+1.2}_{-1.5}$ $R_{\odot}$\\
 Companion mass $M_{1}$ & $9.2^{+1.6}_{-2.2}$ $M_{\odot}$\\
 \hline
 \end{tabular}
\end{table}

\bsp	
\label{lastpage}
\end{document}